\documentstyle[11pt]{article}
\bibliography{plain}
\title{\bf Lewenstein-Sanpera decomposition For Bell Decomposable
States } \vspace{20mm}
\author{
S. J. Akhtarshenas  $^{a,b,c}$ \thanks{E-mail:akhtarshenas@tabrizu.ac.ir}
 , M. A. Jafarizadeh$^{a,b,c}$ \thanks{E-mail:jafarizadeh@tabrizu.ac.ir}
\\
\\
$^a${\small Department of Theoretical Physics and Astrophysics,
Tabriz University, Tabriz 51664, Iran.} \\
$^b${\small Institute for Studies in Theoretical Physics and Mathematics,
 Tehran 19395-1795, Iran.} \\
$^c${\small Research Institute for Fundamental Sciences, Tabriz
51664, Iran.}} \pagebreak

 \newtheorem{thm}{Theorem}
 
 \newtheorem{lem}[thm]{Lemma}

\begin{document}
\maketitle \vspace{15mm}
\newpage
\begin{abstract}
We propose a simple geometrical approach for finding the
Lewenstein-Sanpera decomposition of Bell decomposable states of
$2\otimes 2$ quantum systems. We show that in these systems, the
weight of the pure entangled part in the decomposition is equal to
the concurrence of the state. It is also shown that the optimized
separable part of L-S decomposition  minimizes the von Neumann
relative entropy. We also obtain the decomposition for a class of
mixed states by using some LQCC actions. It is also shown that for
these states the average concurrence of L-S decomposition is equal
to their concurrence.

{\bf Keywords: Quantum entanglement, Bell decomposable states,
Lewenstein-Sanpera decomposition, Concurrence}

{\bf PACs Index: 03.65.U }
\end{abstract}
\pagebreak

\vspace{7cm}

\section{Introduction}
Entanglement is one of the most striking features of quantum
mechanics \cite{EPR,shcro}. The non local character of an
entangled system is usually manifested in quantum correlations
between non interacting subsystems provided that they had only
interaction in the past. A bipartite mixed state is said to be
separable (non entangled) if it can be written as a convex
combination of pure states
\begin{equation}
\rho=\sum_{i}p_{i}\left|\phi_{i}^{A}\right>\left<\phi_{i}^{A}\right|
\otimes\left|\psi_{i}^{B}\right>\left<\psi_{i}^{B}\right|,
\end{equation}
where $\left|\phi_{i}^{A}\right>$ and $\left|\psi_{i}^{B}\right>$
are pure states of subsystems $A$ and $B$, respectively. In the
case of pure states it is easy to check whether a given state is,
or is not entangled. Entangled pure states do always violate Bell
inequalities \cite{Bell}. For mixed states, however, the
statistical properties of the mixture can hide the quantum
correlations embodied in the system, making thus the distinction
between separable and entangled states enormously difficult.

In the pioneering parer \cite{LS}, a very interesting description
of entanglement was achieved by defining the best separable
approximation (BSA) of a mixed state. In the case of 2-qubit
system, it consists of a decomposition of the state into a linear
combination of mixed separable part and a pure entangled one. In
this way, the whole non-separability properties are concentrated
in the pure part. It also provides a natural measure of
entanglement given by the entanglement of the pure part (well
defined for pure states) multiplied by the weight of the pure part
in the composition.

In the Ref. \cite{LS}, the numerical method for finding the BSA
has been reported. Also in $2\otimes2$ systems some analytical
results for special states were found in \cite{englert}. An
attempt to generalize the results of Ref. \cite{LS} is made in
\cite{karnas}.

In \cite{kus} an algebraic approach to find BSA of a 2-qubit state
is attempted. They have also showed that the weight of the
entangled part in the decomposition is equal to the concurrence of
the state.

In this paper we consider Bell decomposable (BD) states. We
provide a simple geometrical approach and give an analytical
expression for L-S decomposition, where our results are in
agreement with those reported in \cite{LS,englert}. Our method to
find L-S decomposition is geometrically intuitive.  We also see
that the weight of the entangled part in the decomposition is
equal to the concurrence of the state. It is also shown that
separable state optimizing L-S decomposition,  minimizes the von
Neumann relative entropy introduced in \cite{ved1,ved2} as a
measure of entanglement. Starting from BD states, we perform local
quantum operations and classical communications (LQCC) and find
L-S decomposition for a generic two qubit system. We prove that
for some special LQCC the obtained decomposition is optimal. It is
also shown that for these cases average concurrence of the
decomposition is equal to concurrence.

The paper is organized as follows. In section 2 we review BD
states and present a perspective of their geometry. L-S
decomposition of these states is obtained in section 3 via a
geometric approach. We prove that this decomposition is optimal.
Relation between L-S decomposition and relative entropy is
discussed  in section 4. It is shown that BSA also minimize von
Neumann relative entropy. Effect of LQCC on L-S decomposition is
studied in section 5. The paper is ended with a brief conclusion.

\section{Bell Decomposable States} In
this section we briefly review Bell decomposable (BD) states and
some of their properties. A BD state is defined by
\begin{equation}
\rho=\sum_{i=1}^{4}p_{i}\left|\psi_i\right>\left<\psi_i\right|,\quad\quad
0\leq p_i\leq 1,\quad \sum_{i=1}^{4}p_i=1,
 \label{BDS1}
\end{equation}
where $\left|\psi_i\right>$ is Bell state given by
\begin{eqnarray}
\label{BS1} \left|\psi_1\right>=\left|\phi^{+}\right>
=\frac{1}{\sqrt{2}}(\left|\uparrow\uparrow\right>+\left|
\downarrow\downarrow\right>), \\
\label{BS2}\left|\psi_2\right>=\left|\phi^{-}\right>
=\frac{1}{\sqrt{2}}(\left|\uparrow\uparrow\right>-\left|
\downarrow\downarrow\right>), \\
\label{BS3}\left|\psi_3\right>=\left|\psi^{+}\right>
=\frac{1}{\sqrt{2}}(\left|\uparrow\downarrow\right>+\left|
\downarrow\uparrow\right>), \\
\label{BS4}\left|\psi_4\right>=\left|\psi^{-}\right>
=\frac{1}{\sqrt{2}}(\left|\uparrow\downarrow\right>-\left|
\downarrow\uparrow\right>).
\end{eqnarray}
In terms of Pauli's matrices, $\rho$ can be written as

\begin{equation}
\rho=\frac{1}{4}(I\otimes I+\sum_{i=1}^{3}
t_i\sigma_{i}\otimes\sigma_{i}), \label{BDS2}
\end{equation}
where

\begin{equation}\label{t-p}
\begin{array}{rl}
t_1=&p_1-p_2+p_3-p_4,  \\
t_2=&-p_1+p_2+p_3-p_4, \\
t_3=&p_1+p_2-p_3-p_4.
\end{array}
\end{equation}

From positivity of $\rho$ we get
\begin{equation}\label{T1}
\begin{array}{rl}
1+t_1-t_2+t_3\geq & 0,  \\
1-t_1+t_2+t_3\geq & 0,  \\
1+t_1+t_2-t_3\geq & 0,  \\
1-t_1-t_2-t_3\geq & 0.
\end{array}
\end{equation}
These equations form a tetrahedral  with its vertices located at
$(1,-1,1)$, $(-1,1,1)$, $(1,1,-1)$, $(-1,-1,-1)$ \cite{horo2}. In
fact these vertices are Bell states given in Eqs. (\ref{BS1}) to
(\ref{BS4}), respectively.

According to the Peres and Horodecki's condition for separability
\cite{peres,horo1}, a 2-qubit state is separable if and only if
its partial transpose is positive. This implies that $\rho$ given
in Eq. (\ref{BDS2}) is separable if and only if $t_i$s satisfy
Eqs. (\ref{T1}) and
\begin{equation}\label{T2}
\begin{array}{rl}
1+t_1+t_2+t_3\geq & 0,  \\
1-t_1-t_2+t_3\geq & 0,  \\
1+t_1-t_2-t_3\geq & 0,  \\
1-t_1+t_2-t_3\geq & 0.
\end{array}
\end{equation}

Inequalities (\ref{T1}) and (\ref{T2}) form an octahedral with its
vertices located at $O_1^{\pm}=(\pm 1,0,0)$, $O_2^{\pm}=(0,\pm
1,0)$ and $O_3^{\pm}=(0,0,\pm 1)$. Hence, tetrahedral of Eqs.
(\ref{T1}) is divided into five regions. Central regions, defined
by octahedral, are separable states. There are also four smaller
equivalent tetrahedral corresponding to entangled states. Each
tetrahedral takes one Bell state as one of its vertices. Three
other vertices of each tetrahedral form a triangle which is its
common face with the octahedral (See Fig. 1).

\section{Lewenstein-Sanpera decomposition}
According to Lewenstein-Sanpera decomposition \cite{LS}, any
2-qubit density matrix $\rho$ can be written as
\begin{equation}\label{LSD}
\rho=\lambda\rho_{sep}+(1-\lambda)\left|\psi\right>\left<\psi\right|,\quad\quad
\lambda\in[0,1],
\end{equation}
where $\rho_{sep}$ is a separable density matrix and
$\left|\psi\right>$ is a pure entangled state. The L-S
decomposition of a given density matrix $\rho$ is not unique and,
in general, there is a continuum set of L-S decomposition to
choose from. The optimal decomposition is unique in which
$\lambda$ is maximal, and
\begin{equation}\label{LSDopt}
\rho=\lambda^{(opt)}\rho_{sep}^{(opt)}
+(1-\lambda^{(opt)})\left|\psi^{(opt)}\right>\left<\psi^{(opt)}\right|\;,
\quad\quad
\lambda^{(opt)}\in[0,1].
\end{equation}
 All other decomposition of the form
 $\rho={\tilde \lambda}{\tilde \rho}_{sep}
 +(1-{\tilde \lambda})\left|{\tilde \psi}\right>\left<{\tilde \psi}\right|$
 , with ${\tilde \lambda}\in[0,1]$ such that ${\tilde
 \rho}\neq\rho^{(opt)}$ necessarily implies that ${\tilde
 \lambda}<\lambda^{(opt)}$ \cite{LS}.

 In the following, we will
refer to Eq. (\ref{LSD}) as the optimal decomposition of $\rho$.
The separable part $\rho_{sep}$ is called the best separable
approximation (BSA) of $\rho$, and $\lambda$ is its separability.

Here in this section we obtain L-S decomposition for Bell
decomposable states via a geometrical approach. Our results are in
agreement with those reported by \cite{LS,englert}. In addition we
present an explicit form for $\rho_{sep}$ and show that, pure
entangled state $\left|\psi\right>$ is Bell state which $\rho$
belongs to its entangled tetrahedral. For simplicity, we show in
Fig. 2 entangled tetrahedral corresponding to singlet state
(\ref{BS4}).

Suppose $\rho$ is an entangled state parameterized as
$\overrightarrow{t}=(t_1,t_2,t_3)$. We connect vertex $p$, which
denotes singlet state, to point $\overrightarrow{t}$ and extend it
to cut separable surface $O_1^{-}O_2^{-}O_3^{-}$ at
$\overrightarrow{t^{\prime}}$ corresponding to separable state
$\rho_s$, where this line can defined by Eqs.
$(1+t_2)(x_1-t_1)-(1+t_1)(x_2-t_2)=0$ and
$(1+t_3)(x_2-t_2)-(1+t_2)(x_3-t_3)=0$. It can be easily seen that
this line cuts plane $O_1^{-}O_2^{-}O_3^{-}$, defined by
$x_1+x_2+x_3+1=0$, at point
$\overrightarrow{t^\prime}=(t^{\prime}_1,t^{\prime}_2,t^{\prime}_3)$

\[t^\prime_{1}= \frac{-1+t_1-t_2-t_3}{3+t_1+t_2+t_3},\]
\vspace{-7mm}\begin{equation}\label{tp1}
t^\prime_{2}=\frac{-1-t_1+t_2-t_3}{3+t_1+t_2+t_3},
\end{equation}
\[t^\prime_{3}=\frac{-1-t_1-t_2+t_3}{3+t_1+t_2+t_3}.\]
Using Eq. (\ref{t-p}) it is straightforward to obtain coordinates
of $\rho_{s}$ in terms of parameters $p_{i}$ as
\begin{eqnarray}\label{pp1}
p_i^{\prime}=\frac{p_i}{2(1-p_{4})}\quad\quad{\mbox for}\quad
i=1,2,3  \qquad\mbox{and}\quad p_4^{\prime}=\frac{1}{2}.
\end{eqnarray}

Now, using the Eqs. (\ref{BDS2}), (\ref{tp1}) we can write
explicit form for separable state $\rho_s$
\begin{equation}\label{rhosep}
\rho_s=\frac{1}{2(3+t_1+t_2+t_3)}\left(\begin{array}{cccc}
 1+t_3& 0 & 0 & t_1-t_2 \\
 0 & 2+t_1+t_2 & -1-t_3 & 0 \\
 0 & -1-t_3 & 2+t_1+t_2 & 0 \\
 t_1-t_2 & 0 & 0 & 1+t_3
 \end{array}\right).
\end{equation}
By convexity we can write $\rho$ as convex sum of $\rho_s$ and
projector $\left|\psi^{-}\right>\left<\psi^{-}\right|$
\begin{equation}\label{LSD1}
\rho=\lambda\rho_{s}+(1-\lambda)\left|\psi^{-}\right>\left<\psi^{-}\right|.
\end{equation}

Using Eqs. (\ref{tp1}) and (\ref{LSD1}) we obtain
\begin{equation}\label{lambda}
\lambda=\frac{|pt|}{|pt^{\prime}|}=\frac{3+t_1+t_2+t_3}{2}=1-C,
\end{equation}
where $|pt|$ and $|pt^\prime|$ are distances between points $p$,
$t$ and also $p$, $t^\prime$ respectively, and $C$ is concurrence
of $\rho$ \cite{woot}.

Obviously Eq. (\ref{lambda}) implies that the entanglement
contribution of singlet state in L-S decomposition of the BD
states is the same as its concurrence. The concurrence of a mixed
state is defined as the minimum of the average concurrence over
all decompositions of the state in terms of pure states
\cite{woot}. This means that for the L-S decomposition given in
(\ref{LSD}) we have $C(\rho)\le (1-\lambda)C(\psi)$. Eq.
(\ref{lambda}) shows that optimal decomposition of BD states
saturate this inequality.

Now, in order to show that $\lambda$ of Eq. (\ref{lambda}) is
maximal and thus the decomposition (\ref{LSD1}) with $\rho_s$
given in Eq. (\ref{rhosep}) is optimal, first we show that
$\rho_s$ can be written in terms of product states. In fact
$\rho_{s}$ of Eq. (\ref{rhosep}) can be written as a convex sum of
three states corresponding to three vertices
$O_1^{-},O_2^{-},O_3^{-}$ of octahedral,
\begin{equation}\label{rhosep123}
\rho_{s}=\lambda_1^{-}\rho_1^{-}+\lambda_2^{-}\rho_2^{-}+
\lambda_3^{-}\rho_3^{-},
\end{equation}
where

\begin{equation}\label{rho123}
\begin{array}{rl}
\rho_1^{-}= &
\frac{1}{2}(\left|x_{+}\right>\left<x_{+}\right|\otimes\left|x_{-}\right>
\left<x_{-}\right|
+\left|x_{-}\right>\left<x_{-}\right|\otimes\left|x_{+}\right>
\left<x_{+}\right|),
\\
\rho_2^{-}= &
\frac{1}{2}(\left|y_{+}\right>\left<y_{+}\right|\otimes\left|y_{-}\right>
\left<y_{-}\right|
+\left|y_{-}\right>\left<y_{-}\right|\otimes\left|y_{+}\right>\left<y_{+}
\right|),
\\
\rho_3^{-}= &
\frac{1}{2}(\left|z_{+}\right>\left<z_{+}\right|\otimes\left|z_{-}\right>
\left<z_{-}\right|
+\left|z_{-}\right>\left<z_{-}\right|\otimes\left|z_{+}\right>\left<z_{+}
\right|),
\end{array}
\end{equation}
and $\left|x_{\pm}\right>$, $\left|y_{\pm}\right>$ and
$\left|z_{\pm}\right>$ are eigenstates  corresponding to
eigenvalues $\pm1$ of $\sigma_{x}$, $\sigma_{y}$ and $\sigma_{z}$,
respectively. Considering the fact that for any point interior to
equilateral triangles, the  sum of the orthogonal distances of the
point to three edges is equal to the height of triangle, for
triangle $O_1^{-}O_2^{-}O_3^{-}$ we have $h_1+h_2+h_3=\sqrt{3/2}$,
where $h_1$, $h_2$ and $h_3$ are orthogonal distances from
corresponding edges (See Fig. 3). After straightforward
calculation we get
\[ h_i=-\sqrt{\frac{3}{2}}t_i^\prime\;, \]
 where $t_i^\prime$, ($i=1,2,3$) are coordinates of $\rho_{s}$ given by
  (\ref{tp1}).

Taking into account the fact that $\lambda_{i}^{-}$ is
proportional to $h_i$ and
$\lambda=\lambda_1^{-}+\lambda_2^{-}+\lambda_3^{-}$, we get

\begin{equation}\label{lambda123}
\begin{array}{rl}
\lambda_1^{-}= & \frac{1}{2}(1-t_1+t_2+t_3), \\
\lambda_2^{-}= & \frac{1}{2}(1+t_1-t_2+t_3), \\
\lambda_3^{-}= & \frac{1}{2}(1+t_1+t_2-t_3).
\end{array}
\end{equation}

The same is true for other Bell decomposable states belonging to
other maximally entangled tetrahedral.

Using above results we rewrite $\rho$ given in Eq. (\ref{LSD1}) in
terms of product states and pure entangled state

\begin{equation}\label{rhosep2}
\rho=\sum_{\alpha=1}^{6}\Lambda_{\alpha}\left|e_{\alpha}
,f_{\alpha}\right>
\left<e_{\alpha},f_{\alpha}\right|+\left|\psi^{-}\right>\left<\psi^{-}\right|,
\end{equation}
where $\left|e_{\alpha} ,f_{\alpha}\right>$, $\alpha=1,2,...,6$
are product states defined by

\begin{equation}\label{pro1-6}
\begin{array}{lcl}
\left|e_{1} ,f_{1}\right>=
\left|x_{+}\right>\otimes\left|x_{-}\right> &,\quad &
\left|e_{2},f_{2}\right>=
\left|x_{-}\right>\otimes\left|x_{+}\right>, \\
\left|e_{3},f_{3}\right>=
\left|y_{+}\right>\otimes\left|y_{-}\right> &,\quad &
\left|e_{4},f_{4}\right>=
\left|y_{-}\right>\otimes\left|y_{+}\right>, \\
\left|e_{5},f_{5}\right>=
\left|z_{+}\right>\otimes\left|z_{-}\right> &,\quad &
\left|e_{6},f_{6}\right>=
\left|z_{+}\right>\otimes\left|z_{-}\right>,
\end{array}
\end{equation}
and $\Lambda_{\alpha}$, $\alpha=1,2,...6$ are given by
\begin{eqnarray}
\Lambda_1=\Lambda_2=\frac{\lambda_1^{-}}{2},  \\
\Lambda_3=\Lambda_4=\frac{\lambda_2^{-}}{2},  \\
\Lambda_5=\Lambda_6=\frac{\lambda_3^{-}}{2}.
\end{eqnarray}

Now with this notation, we are in position to prove that
decomposition (\ref{LSD1}) with $\lambda$ given in (\ref{lambda})
is optimal. To do this we show that all coefficients of product
states appeared in (\ref{rhosep2}) are maximal. According to
\cite{LS} maximizing all the pairs
$(\Lambda_{\alpha},\Lambda_{\beta})$ with respect to
$\rho_{\alpha\beta}=
\rho-\sum_{\alpha^\prime\neq\alpha,\beta}\Lambda_{\alpha^\prime}
P_{\alpha^\prime}$
and $(P_{\alpha},P_{\beta})$ is a necessary and sufficient
condition to subtract the maximal separable matrix
$\rho_{s}^{\alpha}$  from $\rho$, where for the sake of
self-containty   we quote the theorem 2 and the related lemmas of
reference \cite{LS} below.

\begin{thm}\cite{LS}

Given the set $\Lambda_{V}$ of product vectors
$\left|e,f\right>\in {\cal R}(\rho)$ , the matrix
$\rho_s^{\ast}=\sum_{\alpha}\Lambda_{\alpha}P_{\alpha}$ is the
best separable approximation to $\rho$ iff a) all $\Lambda_\alpha$
are maximal with respect to
$\rho_{\alpha}=\rho-\sum_{\alpha^\prime\neq\alpha}\Lambda_{\alpha^\prime}
P_{\alpha^{\prime}}$
and the projector $P_{\alpha}$ ; b) all pairs
$(\Lambda_{\alpha},\Lambda_{\beta})$ are maximal with respect to
$\rho_{\alpha\beta}=\rho-\sum_{\alpha^\prime\neq\alpha,\beta}
\Lambda_{\alpha^\prime}P_{\alpha^\prime}$,
and the projectors $(P_{\alpha},P_{\beta})$.
\end{thm}

\begin{lem}\cite{LS}
$\Lambda$ is maximal with respect to $\rho$ and
$P=\left|\psi\right>\left<\psi\right|$ iff a) if
$\left|\psi\right>\not\in {\cal R}(\rho)$ then $\Lambda=0$, and b)
if $\left|\psi\right>\in {\cal R}(\rho)$ then
$\Lambda=\left<\psi\right|\rho^{-1}\left|\psi\right>^{-1}>0$.
\end{lem}

\begin{lem}\cite{LS}
A pair $(\Lambda_1,\Lambda_2)$ is maximal with respect to $\rho$
and a pair of projectors $(P_1,P2)$ iff: a) if
$\left|\psi_1\right>$, $\left|\psi_2\right>$ do not belong to
${\cal R}(\rho)$ then $\Lambda_1=\Lambda_2=0$; b) if
$\left|\psi_1\right>$ does not belong, while
$\left|\psi_2\right>\in{\cal R}(\rho)$ then $\Lambda_1=0$,
$\Lambda_2=\left<\psi_2\right|\rho^{-1}\left|\psi_2\right>^{-1}$;
c) if $\left|\psi_1\right>$, $\left|\psi_2\right>\in {\cal
R}(\rho)$ and $\left<\psi_1\right|\rho^{-1}\left|\psi_2\right>=0$
then
$\Lambda_i=\left<\psi_i\right|\rho^{-1}\left|\psi_i\right>^{-1}$,
$i=1,2$; d) finally, if $\left|\psi_1\right>,
\left|\psi_2\right>\in {\cal R}(\rho)$ and
$\left<\psi_1\right|\rho^{-1}\left|\psi_2\right>\neq 0$ then

\begin{equation}\label{Lambda12}
\begin{array}{lr}
\Lambda_1= &(\left<\psi_2\right|\rho^{-1}\left|\psi_2\right>
-\mid\left<\psi_1\right|\rho^{-1}\left|\psi_2\right>\mid)/D, \\
\Lambda_2= &(\left<\psi_1\right|\rho^{-1}\left|\psi_1\right>
-\mid\left<\psi_1\right|\rho^{-1}\left|\psi_2\right>\mid)/D,
\end{array}
\end{equation}
where
$D=\left<\psi_1\right|\rho^{-1}\left|\psi_1\right>\left<\psi_2\right|
\rho^{-1}\left|\psi_2\right>
-\mid\left<\psi_1\right|\rho_{-1}\left|\psi_2\right>\mid^2$.
\end{lem}

First, we show that $\Lambda_{\alpha}$s are maximal with respect
to $\rho_{\alpha}$ and $P_{\alpha}$.

Matrices  $\rho_{\alpha}=\rho-\sum_{\alpha^{\prime}\neq
\alpha}^{6}=\Lambda_{\alpha}
P_{\alpha}+(1-\lambda)\left|\psi^{-}\right>\left<\psi^{-}\right|$
with
$P_{\alpha}=\left|e_{\alpha},f_{\alpha}\right>\left<e_{\alpha},
f_{\alpha}\right|$
 for ($i=1,2,...,6$) have two zero eigenvalues and two non zero
eigenvalues. In Bell basis its kernel and range are separated.
After restriction to its range, it is straightforward to evaluate
$\rho_i^{-1}$ and we find that
$\left<e_i,f_i\right|\rho_i^{-1}\left|e_i,f_i\right>=1/\Lambda_i$.

In order to prove that the pair $(\Lambda_\alpha,\Lambda_\beta)$
are maximal with respect to $\rho_{\alpha\beta}$ and the pair of
projectors $(P_\alpha,P_\beta)$, we proceed as follows:

a) Matrices
$\rho_{i,i+1}=\Lambda_iP_i+\Lambda_{i+1}P_{i+1}+(1-\lambda)
\left|\psi^{-}\right>\left<\psi^{-}\right|$
 for $(i=1,3,5)$ have a two dimensional range . In Bell basis its
range and kernel are separated and one can obtain
$\left<e_i,f_i\right|\rho^{-1}_{i,i+1}\left|e_i,f_i\right>=
(\Lambda_{i+1}+(1-\lambda))/\Gamma_i,
\;\left<e_{i+1},f_{i+1}\right|\rho^{-1}_{i,i+1}\left|e_{i+1},f_{i+1}\right>
=(\Lambda_{i}+(1-\lambda))/\Gamma_i$ and
$\left<e_i,f_i\right|\rho^{-1}_{i,i+1}\left|e_{i+1},f_{i+1}\right>
=(1-\lambda))/(2\Gamma_i)$,
where $\Gamma_i=\Lambda_{i}\Lambda_{i+1}+\frac{1}{2}(1-\lambda)$.
Using the above results together with Eqs. (\ref{Lambda12}) we
obtain the maximality of pair $(\Lambda_i,\Lambda_{i+1})$ with
respect to $\rho_{i,i+1}$ and the pair of projectors
$(P_i,P_{i+1})$ for $i=1,3$ and $5$.

b) For other possibility of $\alpha$ and $\beta$, matrices
$\rho_{\alpha\beta}=\Lambda_{\alpha}P_{\alpha}+\Lambda_{\beta}P_{\beta}
+(1-\lambda)\left|\psi^{-}\right>\left<\psi^{-}\right|$  have rank
3 . Using the Bell basis we can evaluate $\rho^{-1}$ and we find
that
$\left<e_\alpha,f_{\alpha}\right|\rho^{-1}\left|e_\beta,f_{\beta}\right>=0$
for $\alpha\neq\beta$,
$\left<e_\alpha,f_{\alpha}\right|\rho^{-1}\left|e_{\alpha},f_{\alpha}\right>
=1/\Lambda_{\alpha}$. This completes the proof that
$\Lambda_\alpha$ of Eq. (\ref{lambda123})  are maximal and
decomposition (\ref{LSD1}) is optimal.

Also it is worth to note that the decomposition (\ref{LSD1})
satisfies conditions for BSA of Ref. \cite{kus}. According to
theorem 1 of Ref. \cite{kus}, decomposition given in (\ref{LSD1})
is the optimal decomposition if and only if:
rank$(\rho_s^{T_B})=3$, i.e.
$\exists_{\left|\phi\right>}\,\,\rho_s^{T_{B}}\left|\phi\right>=0$,
and either
\begin{equation}
\begin{array}{rl}
\mbox{(i)} &
\exists_{\alpha>0}\left(\left|\phi\right>\left<\phi\right|\right)^{T_{B}}
\left|\psi\right>
=-\alpha\,\left|\psi\right>,\quad \mbox{or} \\ \mbox{(ii)}
&\mbox{rank}(\rho_s)=3,\, i.e.\,\exists_{\tilde
{\left|\phi\right>}} \,\,\rho_s\tilde{
\left|\phi\right>}=0,\,\mbox{and}\,\,
\exists_{\alpha,\nu\ge0}\,\,\left(\nu\tilde{
 \left|\phi\right>}\tilde{\left<\phi\right|}
+\left(\left|\phi\right>\left<\phi\right|\right)^{T_{B}}\right)
\left|\psi\right>=
-\alpha\left|\psi\right>.
\end{array}
\end{equation}
It is now straightforward  to see that $\rho_{s}^{T_{B}}$ has
three non vanishing eigenvalues, that is, its rank is 3. Its one
dimensional kernel is along the Bell state $\left|\psi_{1}\right>$
given in Eq. (\ref{BS1}). Actually the density matrices
corresponding to the interior of tetrahedral satisfy condition (i)
while those at its boundary satisfy condition (ii), respectively.

\section{Relative entropy of entanglement and L-S decomposition}
Vedral et al. in \cite{ved1,ved2} introduced a class of distance
measures suitable for entanglement measures. According to their
methods, entanglement measure for a given state $\rho$ is defined
as
\begin{equation}\label{D}
E(\rho)=\min_{\sigma\in {\mathcal D}} D(\rho\parallel\sigma),
\end{equation}
where $D$ is any measure of distance (not necessarily a metric)
between two density matrix $\rho$ and $\sigma$, and ${\mathcal D}$
is the set of all separable states. They have also shown that von
Neumann relative entropy defined by
\begin{equation}\label{RE}
S(\rho\parallel\sigma)=tr\{\rho\ln \frac{\rho}{\sigma}\},
\end{equation}
satisfies three conditions that a good measure of entanglement
must satisfy \cite{ved1}. Here, we would like to emphasis that
$\rho_{s}$ given in Eq. (\ref{rhosep}) minimizes von Neumann
relative entropy given in (\ref{RE}). Authors in \cite{ved1} have
shown that for BD states given in Eq. (\ref{BDS1}), separable
state $\sigma$ that minimize relative entropy is
\begin{eqnarray}\label{pp1}
p_i^{\prime}=\frac{p_i}{2(1-p_{4})}\quad\quad{\mbox for}\quad
i=1,2,3  \qquad\mbox{and}\quad p_4^{\prime}=\frac{1}{2}.
\end{eqnarray}

It is worth to note that the above equation is the same as Eq.
(\ref{pp1}), that is, separable state optimizing L-S decomposition
 minimizes von Neumann relative entropy, too.

\section{L-S decomposition under LQCC}
In this section we study the behavior of L-S decomposition under
local quantum operations and classical communications (LQCC). A
general LQCC is defined by \cite{lind,kent}
\begin{equation}\label{lqcc}
\rho^{\prime}=\frac{(A\otimes B)\rho(A\otimes
B)^{\dag}}{tr((A\otimes B)\rho(A\otimes B)^{\dag})},
\end{equation}
where operators $A$ and $B$ can be written as
\begin{equation}
A\otimes B=U_{A}\,f^{\mu,a,{\bf m}}\otimes U_{B}\,f^{\nu,b,{\bf
n}},
\end{equation}
where $U_{A}$ and $U_{B}$ are unitary operators acting on
subsystems $A$ and $B$, respectively and the filtration  $f$
defined by
\begin{equation}\label{filt}
\begin{array}{rl}
f^{\mu,a,{\bf m}}= & \mu(I_2 + a\,{\bf m}.{\bf \sigma}), \\
f^{\nu,b,{\bf n}}= & \nu(I_2 + b\,{\bf n}.{\bf \sigma}).
\end{array}
\end{equation}
As it is shown in Refs. \cite{lind,kent}, the concurrence of the
state $\rho$ transforms under LQCC of the form given in Eq.
(\ref{lqcc}) as
\begin{equation}\label{conlqcc}
C(\rho^{\prime})=\frac{\mu^2\,\nu^2(1-a^2)(1-b^2)}{tr((A\otimes
B)\rho(A\otimes B)^{\dag})}\,C(\rho).
\end{equation}

Performing LQCC  on L-S decomposition of BD states we get

\begin{equation}\label{lqcclsd}
\rho^{\prime}= \frac{(A\otimes B)\rho(A\otimes
B)^{\dag}}{tr((A\otimes B)\rho(A\otimes B)^{\dag})}=
\lambda^{\prime}
\rho_s^{\prime}+(1-\lambda^{\prime})\left|\psi^{\prime}\right>
\left<\psi^{\prime}\right|,
\end{equation}
with $\rho_s^{\prime}$ and $\left|\psi^{\prime}\right>$ defined as
\begin{equation}\label{lqccrhos}
\rho_s^{\prime}=\frac{(A\otimes B)\rho_s(A\otimes
B)^{\dag}}{tr((A\otimes B)\rho_s(A\otimes B)^{\dag})},
\end{equation}
\begin{equation}
\left|\psi^{\prime}\right> =\frac{(A\otimes
B)\left|\psi^{-}\right>}{\sqrt{\left<\psi^{-}\right|(AA^{\dag}\otimes
BB^{\dag})\left|\psi^{-}\right>}},
\end{equation}
respectively, and $\lambda^{\prime}$ is
\begin{equation}\label{lqcclam}
\lambda^{\prime}=\frac{tr((A\otimes B)\rho_s(A\otimes
B)^{\dag})}{tr((A\otimes B)\rho(A\otimes B)^{\dag})}\,\lambda.
\end{equation}
Using Eq. (\ref{lqcclam}), we get for  the weight of entangled
part in the decomposition  (\ref{lqcclsd})
\begin{equation}
(1-\lambda^{\prime})=\frac{\left<\psi^{-}\right|(AA^{\dag}\otimes
BB^{\dag})\left|\psi^{-}\right>}{tr((A\otimes B)\rho(A\otimes
B)^{\dag})}\,(1-\lambda).
\end{equation}
Now we can easily  evaluate the average concurrence of
$\rho^{\prime}$ in the L-S decomposition given in (\ref{lqcclsd})
\begin{equation}
(1-\lambda^{\prime})C\left(\left|\psi^{\prime}\right>\right)=
\frac{\mu^2\,\nu^2(1-a^2)(1-b^2)}{tr((A\otimes B)\rho(A\otimes
B)^{\dag})}\,(1-\lambda)C(\left|\psi\right>),
\end{equation}
where, by Comparing the above equation with Eq. (\ref{conlqcc}) we
see that $(1-\lambda)C(\left|\psi\right>)$ (the average
concurrence in the L-S decomposition) transforms  like concurrence
under LQCC.

In order to prove that the decomposition  (\ref{lqcclsd}) is the
optimal one, we rewrite $\rho_s$ in terms of the product states
given in Eq. (\ref{pro1-6})
\begin{equation}\label{rhosep3}
\rho_s=\sum_{\alpha=1}^{6}\Lambda_{\alpha}\left|e_{\alpha}
,f_{\alpha}\right> \left<e_{\alpha},f_{\alpha}\right|.
\end{equation}
Now, performing LQCC action we get
\begin{equation}\label{rhosep3}
\rho_s^{\prime}=\sum_{\alpha=1}^{6}\Lambda_{\alpha}^{\prime}
\left|e_{\alpha}^{\prime}
,f_{\alpha}^{\prime}\right>
\left<e_{\alpha}^{\prime},f_{\alpha}^{\prime}\right|,
\end{equation}
where
\begin{equation}\label{lqccpro}
\left|e_{\alpha}^{\prime} ,f_{\alpha}^{\prime}\right>=
\frac{(A\otimes B)\left|e_{\alpha} ,f_{\alpha}\right>}
{\sqrt{t\left(P_{\alpha}\right )}} ,\qquad
t\left(P_{\alpha}\right)=\left<e_{\alpha},f_{\alpha}\right|(AA^{\dag}\otimes
BB^{\dag})\left|e_{\alpha} ,f_{\alpha}\right>,
\end{equation}
with $P_\alpha=\left|e_{\alpha}
,f_{\alpha}\right>\left<e_{\alpha},f_{\alpha}\right|$ and
\begin{equation}
\Lambda_\alpha^{\prime}=\frac{\left<e_{\alpha},f_{\alpha}
\right|(AA^{\dag}\otimes
BB^{\dag})\left|e_{\alpha} ,f_{\alpha}\right>}{tr((A\otimes
B)\rho(A\otimes B)^{\dag})}\,\Lambda_{\alpha}.
\end{equation}

First we show that $\Lambda_{\alpha}^{\prime}$s are maximal with
respect to $\rho_{\alpha}^{\prime}$ and the projector
$P_{\alpha}^{\prime}$.

As we see the matrices $\rho_{\alpha}=\Lambda_{\alpha}
P_{\alpha}+(1-\lambda)\left|\psi^{-}\right>\left<\psi^{-}\right|$
for $(\alpha=1,2,...,6)$ transform as
\begin{equation}
\rho_\alpha^\prime=\frac{(A\otimes B)\rho_\alpha(A\otimes
B)^{\dag}}{t(\rho_{\alpha})},\qquad t(\rho_{\alpha})=tr((A\otimes
B)\rho_\alpha(A\otimes B)^{\dag}),
\end{equation}
  under LQCC. Using the fact that LQCC transformations are invertible
\cite{kent,vers1,vers2}, we can evaluate
$\rho_\alpha^{\prime^{-1}}$ as
\begin{equation}
\rho_\alpha^{\prime^{-1}}=t(\rho_{\alpha})\,(A^{\dag}\otimes
B^{\dag})^{-1}\rho_\alpha^{-1}(A\otimes B)^{-1}.
\end{equation}
Using the above equation and Eq. (\ref{lqccpro}) we get
\begin{equation}\label{lamimax}
\left<e_\alpha^\prime,f_\alpha^\prime\right|\rho_\alpha^{\prime^{-1}}
\left|e_\alpha^\prime,f_\alpha^\prime\right>=
\frac{t(\rho)}{t(P_\alpha)}
\left<e_\alpha,f_\alpha\right|\rho_\alpha^{-1}
\left|e_\alpha,f_\alpha\right>=\Lambda_\alpha^{\prime}.
\end{equation}
The Eq. (\ref{lamimax}) shows that $\Lambda_{\alpha}^{\prime}$s
are maximal with respect to $\rho_{\alpha}^{\prime}$ and the
projector $P_{\alpha}^{\prime}$.

In order to prove that the pair
$(\Lambda_{\alpha}^{\prime},\Lambda_{\beta}^{\prime})$ are maximal
with respect to $(\rho_{\alpha}^{\prime},\rho_{\beta}^{\prime})$
and $(P_{\alpha}^{\prime},P_{\beta}^{\prime})$, we proceed as
follows:

a) Matrices $\rho_{i,i+1}=\Lambda_{i} P_{i}+\Lambda_{i+1}
P_{i+1}+(1-\lambda)\left|\psi^{-}\right>\left<\psi^{-}\right|$
transform under LQCC as
\begin{equation}
\rho_{i,i+1}^{\prime}=\frac{(A\otimes B)\rho_{i,i+1}(A\otimes
B)^{\dag}}{t(\rho_{i,i+1})},\qquad t(\rho_{i,i+1})=tr((A\otimes
B)\rho_{i,i+1}(A\otimes B)^{\dag}),
\end{equation}
Using the above equation and invertibility of LQCC we arrive at
the following results
$$
\left<e_i^\prime,f_i^\prime\right|\rho_{i,i+1}^{\prime^{-1}}
\left|e_i^\prime,f_i^\prime\right>=
\frac{t(\rho)}{t(P_i)}\frac{\left(\Lambda_{i+1}
+\frac{1}{2}(1-\lambda)\right)}{\Gamma_i},
$$
\vspace{-5mm}
\begin{equation}\label{lamii1max}
\left<e_{i+1}^\prime,f_{i+1}^\prime\right|\rho_{i,i+1}^{\prime^{-1}}
\left|e_{i+1}^\prime,f_{i+1}^\prime\right>=
\frac{t(\rho)}{t(P_i)}\frac{\left(\Lambda_{i}
+\frac{1}{2}(1-\lambda)\right)}{\Gamma_i},
\end{equation}
$$
\left<e_{i}^\prime,f_{i}^\prime\right|\rho_{i,i+1}^{\prime^{-1}}
\left|e_{i+1}^\prime,f_{i+1}^\prime\right>=
\frac{t(\rho)}{\sqrt{t(P_i)t(P_{i+1})}}\frac{(1-\lambda)}{2\Gamma_i},
$$
where
$\Gamma_{i}=\Lambda_{i}\Lambda_{i+1}+\frac{1}{2}(1-\lambda)
(\Lambda_{i}\Lambda_{i+1})$.
Now using Eqs. (\ref{Lambda12}) we get
$$
\hspace{-60mm}
\frac{\left<e_{i}^\prime,f_{i}^\prime\right|\rho_{i,i+1}^{\prime^{-1}}
\left|e_{i}^\prime,f_{i}^\prime\right>-
\mid\left<e_{i}^\prime,f_{i}^\prime\right|\rho_{i,i+1}^{\prime^{-1}}
\left|e_{i+1}^\prime,f_{i+1}^\prime\right>\mid}{D}=
\Lambda_{i+1}^{\prime}
$$
\begin{equation}\label{lqccd1}
\hspace{81mm}
+(1-\lambda)\frac{\left(t(P_{i+1})-\sqrt{t(P_{i})t(P_{i+1})}
\right)}{t(\rho)},
\end{equation}
$$
\hspace{-50mm}
\frac{\left<e_{i+1}^\prime,f_{i+1}^\prime\right|\rho_{i,i+1}^{\prime^{-1}}
\left|e_{i+1}^\prime,f_{i+1}^\prime\right>-
\mid\left<e_{i}^\prime,f_{i}^\prime\right|\rho_{i,i+1}^{\prime^{-1}}
\left|e_{i+1}^\prime,f_{i+1}^\prime\right>\mid}{D}=\Lambda_{i}^{\prime}
$$
\begin{equation}\label{lqccd2}
\hspace{80mm}
+(1-\lambda)\frac{\left(t(P_{i+1})-\sqrt{t(P_{i})t(P_{i+1})}\right)}{t(\rho)}.
\end{equation}

b) For other  values of  $\alpha$ and $\beta$, matrices
$\rho_{\alpha,\beta}=\Lambda_{\alpha} P_{\alpha}+\Lambda_{\beta}
P_{\beta}+(1-\lambda)\left|\psi^{-}\right>\left<\psi^{-}\right|$
transform under LQCC as
\begin{equation}
\rho_{\alpha,\beta}^{\prime}=\frac{(A\otimes
B)\rho_{\alpha,\beta}(A\otimes
B)^{\dag}}{t(\rho_{\alpha,\beta})},\qquad
t(\rho_{\alpha,\beta})=tr((A\otimes B)\rho_{\alpha,\beta}(A\otimes
B)^{\dag}).
\end{equation}
So with same procedure we can evaluate the following expressions
$$
\left<e_{\alpha}^\prime,f_{\alpha}^\prime
\right|\rho_{\alpha,\beta}^{\prime^{-1}}
\left|e_{\alpha}^\prime,f_{\alpha}^\prime\right>=
\frac{1}{\Lambda_{\alpha}^{\prime}}, $$
\begin{equation}
\left<e_{\beta}^\prime,f_{\beta}^\prime\right|\rho_{\alpha,\beta}^{\prime^{-1}}
\left|e_{\beta}^\prime,f_{\beta}^\prime\right>=
\frac{1}{\Lambda_{\beta}^{\prime}},
\end{equation}
$$
\left<e_{\alpha}^\prime,f_{\alpha}^\prime\right|
\rho_{\alpha,\beta}^{\prime^{-1}}
\left|e_{\beta}^\prime,f_{\beta}^\prime\right>=0, \quad
\mbox{for}\quad \alpha\ne\beta. $$

Eqs. (\ref{lqccd1}) and (\ref{lqccd2}) show that the pair
$(\Lambda_{\alpha}^{\prime},\Lambda_{\beta}^{\prime})$ are maximal
with respect to $\rho_{\alpha,\beta}$ and
$(P_{\alpha}^{\prime},P_{\beta}^{\prime})$ provided that
$t(P_{i})=t(P_{i+1})$. This restricts LQCC to special case of
$A=B$. Under these conditions the decomposition given in Eq.
(\ref{lqcclsd}) is optimal. Actually considering the fact that the
inequality
$C(\rho^{\prime})\le(1-\lambda^{\prime})C(\psi^{\prime})$ is
saturated by decomposition given in (\ref{lqcclsd}) we conjecture
that this decomposition is optimal for a general LQCC, that is,
$\rho_{s}^{\prime}$ is the BSA for $\rho^{\prime}$. This implies
that, in general,  the product states given in Eq. (\ref{lqccpro})
is not the good product ensemble for the best separable part. To
prove  that the decomposition (\ref{lqcclsd}) is  the optimal one
for the whole class of LQCC is still an open problem which is
under investigation.

\section{Conclusion}
We have derived Lewenstein-Sanpera decomposition for Bell
decomposable states from an entirely different approach. We show
that for these systems, the weights of the pure entangled part in
the decomposition is equal to the concurrence of the states.
Optimality of the presented decomposition have been proved by
using the theorems given  in \cite{LS}. It is also shown that the
optimized separable part of L-S decomposition  minimizes the von
Neumann relative entropy. We have also obtained Lewenstein-Sanpera
decomposition for a large class of states obtained from BD states
via some LQCC action. It is shown that for these states the
average concurrence of the decomposition is equal to their
concurrence.

\newpage

\vspace{10mm}

{\Large {\bf Figures Captions}}

\vspace{10mm}

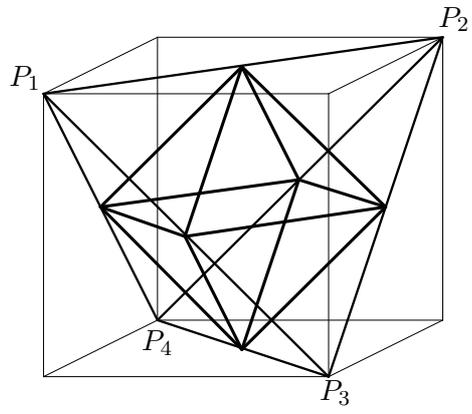
Figure 1: All BD states are defined as points interior to
tetrahedral. Vertices $P_{1}$, $P_{2}$, $P_{3}$ and $P_{4}$ denote
projectors corresponding to Bell states Eqs. (\ref{BS1}) to
(\ref{BS4}), respectively. Octahedral corresponds to separable
states.

\vspace{10mm}

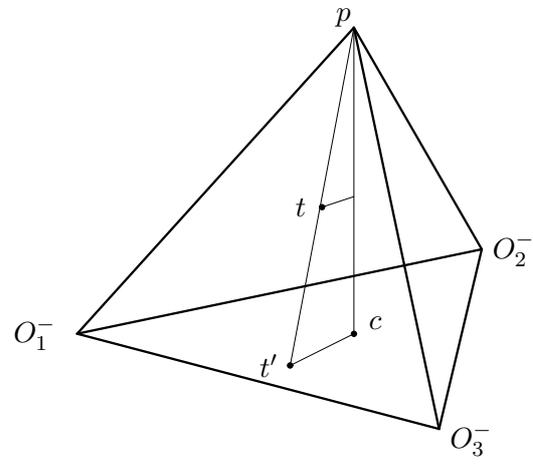
Figure 2: Entangled tetrahedral corresponding to singlet state.
Line $p\,c$ denotes entangled Werner states. Points $t$ and
$t^{\prime}$ correspond to a generic BD state $\rho$ and
associated BSA $\rho_{s}$. Vertex $p$ denote singlet state and
other vertices are defined in Eq. (\ref{rho123}).

\vspace{10mm}

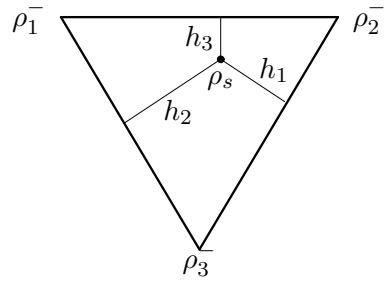
Figure 3: $\rho_{s}$ can be written as a convex combination of
separable states $\rho_{1}^{-}$, $\rho_{2}^{-}$ and
$\rho_{3}^{-}$with weights proportional to $h_{1}$, $h_{2}$ and
$h_{3}$, respectively.

\end{document}